\documentclass{article}
\usepackage{spconf,amsmath,graphicx}
\usepackage{amsmath}
\usepackage{amssymb}
\usepackage{tikz}
\usepackage{wrapfig}
\usepackage{booktabs}
\usepackage{subcaption}

\title{COARSE-FINE SPECTRAL-AWARE DEFORMABLE CONVOLUTION FOR HYPERSPECTRAL IMAGE RECONSTRUCTION}
\name{Jincheng Yang\textsuperscript{1}, Lishun Wang\textsuperscript{2}, Miao Cao\textsuperscript{2}, Huan Wang\textsuperscript{2}, Yinping Zhao\textsuperscript{1}, Xin Yuan\textsuperscript{2}}
\address{\textsuperscript{1}Northwestern Polytechnical University, Xi'an, China\\
            \textsuperscript{2}Westlake University, Hangzhou, China}
\begin{document}
%\ninept
%
\maketitle
\begin{abstract}
We study the inverse problem of Coded Aperture Snapshot Spectral Imaging (CASSI), which captures a spatial-spectral data cube using snapshot 2D measurements and uses algorithms to reconstruct 3D hyperspectral images (HSI). However, current methods based on Convolutional Neural Networks (CNNs) struggle to capture long-range dependencies and non-local similarities. The recently popular Transformer-based methods are poorly deployed on downstream tasks due to the high computational cost caused by self-attention. In this paper, we propose Coarse-Fine Spectral-Aware Deformable Convolution Network (CFSDCN), applying deformable convolutional networks (DCN) to this task for the first time. Considering the sparsity of HSI, we design a deformable convolution module that exploits its deformability to capture long-range dependencies and non-local similarities. In addition, we propose a new spectral information interaction module that considers both coarse-grained and fine-grained spectral similarities. Extensive experiments demonstrate that our CFSDCN significantly outperforms previous state-of-the-art (SOTA) methods on both simulated and real HSI datasets.
\end{abstract}
\section{INTRODUCTION}
\label{sec:INTRODUCTION}
\begin{figure}[t]
    \centering
    \includegraphics[width=\linewidth]{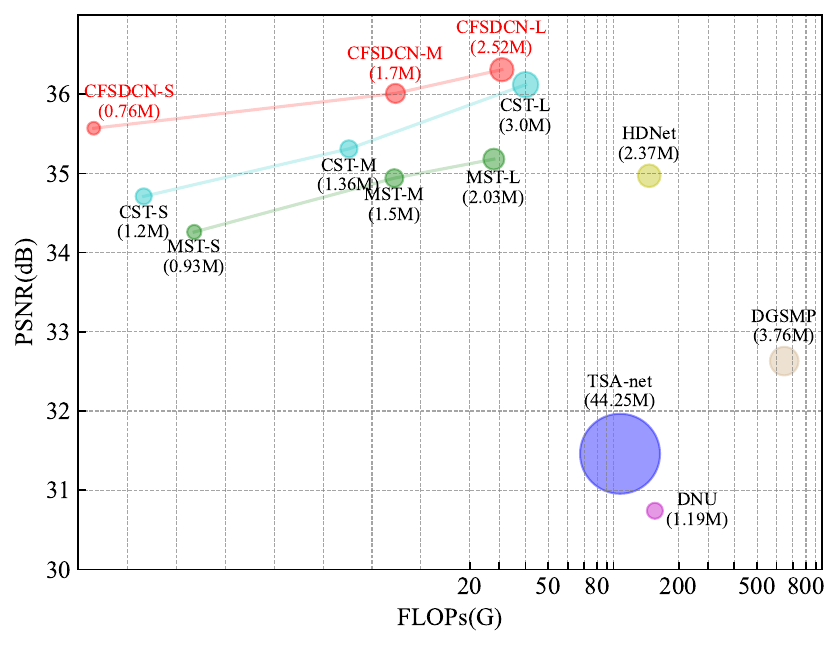}
    \caption{Comparison of PSNR-Params-FLOPs with some previous SOTA methods.The vertical axis represents PSNR, and the horizontal axis represents FLOPs. The size of the scatter points is related to the parameters of the model.}
    \label{fig:first}
\vspace{-4mm}
\end{figure}
Hyperspectral images (HSI) can capture data over a larger number of spectral bands than traditional three-color (RGB) images, thereby storing more comprehensive spectral information. This spectral characteristic enables HSIs to be extensively employed in fields such as image recognition \cite{imgrec01}, object detection \cite{objdetect01}, remote sensing \cite{remotesensing01}, \textit{etc}.

Coded Aperture Snapshot Spectral Imaging (CASSI) captures HSI by using coded apertures and dispersors across different wavelengths to modulate the HSI signal and combine them into two-dimensional compressed measurements, greatly improving frame rates. But it also brings with it the complex task of accurately recovering HSI from compressed measurements.

In recent years, there have been four main solutions to this task. (i) Traditional methods mainly rely on model-based techniques that utilize hand-crafted image priors \cite{TwIST,GAPNet}. They are rooted in strong theoretical foundations and provide interpretability, but their reliance on manual parameter tuning hinders efficiency and better performance. (ii) Plug-and-play (PnP) algorithms \cite{plug04,zheng2021Deep} stand out by seamlessly integrating pre-trained denoising networks within traditional model-based paradigms. However, their reliance on static, immutable networks limits their potential. (iii) The end-to-end (E2E) \cite{TSA-Net,cnnsci03,Miao19ICCV,wang2022snapshot} strategy mainly utilizes convolutional neural networks (CNN) to provide a direct route from measurement to HSI. But they often ignore the nuances of specific imaging systems. (iv) Traditional deep unrolling methods \cite{DNU,DGSMP,meng2023deep} combine iterative optimization with deep neural networks to combine the power of deep learning with the transparency of traditional methods. But they have difficulty solving system-specific degradation problems and HSI long-range dependencies.

With the remarkable success of transformers \cite{Transformer} in language models, Vision Transformers (ViTs) \cite{vit01} have swept through the field of computer vision. They have also achieved outstanding results in HSI reconstruction \cite{MST,RDLUF}. The Multi-head Self-Attention (MSA) module of the Transformer excels in capturing non-local similarities and long-range dependencies, addressing the limitations of CNN-based reconstruction methods. However, the computational requirements of global attention lead to a significant computational load and high memory complexity. Therefore, it faces challenges in being applied to downstream tasks.

To address the aforementioned issues, in this paper, we propose a novel approach, CFSDCN, for HSI reconstruction. The contributions of our work can be summarized as follows:
\begin{itemize}
    \item [i)] We propose a new CNN-based end-to-end method, CFSDCN, for HSI reconstruction. Use deformable convolutions to capture long-range dependencies and non-local similarities of spatially sparse HSI. To the best of my knowledge, this is the first attempt to apply deformable convolution to this task.
    \item[ii)] We propose a novel Spectral Information Perception module composed of convolutions, named CFSAB (Coarse-Fine Spectral-Aware Block). This module, through the use of large kernel and point-wise convolutions, simultaneously captures both coarse-grained and fine -grained spectral similarities.
    \item[iii)] Our model achieves SOTA performance at extremely low computational cost and parameter size in all simulation scenarios of the same scale. Furthermore, our CFSDCN achieves more pleasing results on real datasets.
\end{itemize}

\section{MATHEMATICAL MODEL Of CASSI}
\label{sec:Mathematical Model of CASSI}

\begin{figure}[t]
    \centering
    \includegraphics[width=\linewidth]{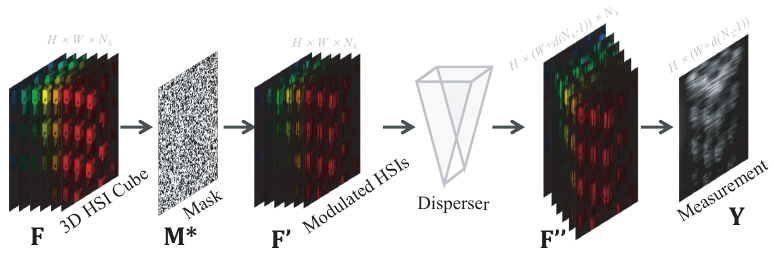}
    \caption{A schematic diagram of CASSI.}
    \label{fig:cassi}
  \vspace{-4mm}
\end{figure}

The schematic of the CASSI system is briefly illustrated in Fig.~\ref{fig:cassi}. The degradation of the CASSI system can be attributed to various factors, including physical masks, dispersive prisms, and 2D imaging sensors. We assume the input 3D HSI data cube to be $\mathbf{F} \in \mathbb{R}^{H \times W \times N_{\lambda}}$, where $H$, $W$, and $N_\lambda$ respectively represent the height, width, and number of wavelengths of the HSI. The physical mask $\mathbf{M^{*}} \in \mathbb{R}^{H \times W}$ serves as the modulator for the HSI signal, representing the modulated image across wavelengths
\begin{equation}
     \mathbf {F'_{n_{\lambda}}} = \mathbf {F_{n_{\lambda}}} \odot \mathbf{M^{*}}, \vspace {-1.2mm} \label {1} 
\end{equation}
where $\mathbf {F' }$ represents the modulated, $n_{\lambda} \in [1,...,N_{\lambda}]$ denotes the $n_{\lambda}^{th}$ wavelength of the modulated image, $\odot$ represents the element-wise product. Therefore, after modulation, the HSI $\mathbf{F'}$ undergoes a displacement during the dispersion process through the disperser, expressed as
\begin{equation}
     \mathbf {F''_{n_{\lambda}}}(u,v,n_{\lambda}) = \mathbf {F'_{n_{\lambda}}}(u,v+d_{n_{\lambda}},n_{\lambda}), \vspace {-1.2mm} \label {2} 
\end{equation}
where $\mathbf{F''} \in \mathbb{R}^{H \times (W + d_{N_{\lambda}})\times N_{\lambda}}$, ($u$, $v$) represents the coordinate system on the detector plane, $d_{n_{\lambda}}$ represents the shifted distance of the $n_{\lambda}^{th}$ wavelength. Finally, the 2D compress measurements $\mathbf{Y} \in \mathbb{R}^{H \times (W + d(N_{\lambda}-1))}$ captured by CASSI can be represented as
\begin{equation}
     \mathbf {Y} =  \textstyle \sum_{n_{\lambda}=1}^{N_{\lambda}} \mathbf {F''_{n_{\lambda}}} \ + \mathbf{G}, \vspace {-1.2mm} \label {3} 
\end{equation}
where $\mathbf {G} \in \mathbb{R}^{H \times (W + d(N_{\lambda}-1))}$ is the imaging noise on the measurement introduced during the system imaging process.
Our task is to restore this 2D measurement $\mathbf{Y}$ captured by CASSI into the 3D spatial-spectral data cube $\mathbf{F}$ we desired.

\section{METHOD}
\label{sec:METHOD}
\subsection{Overall Architecture}
\begin{figure*}[t]
  \centering
  \includegraphics[width=\linewidth]{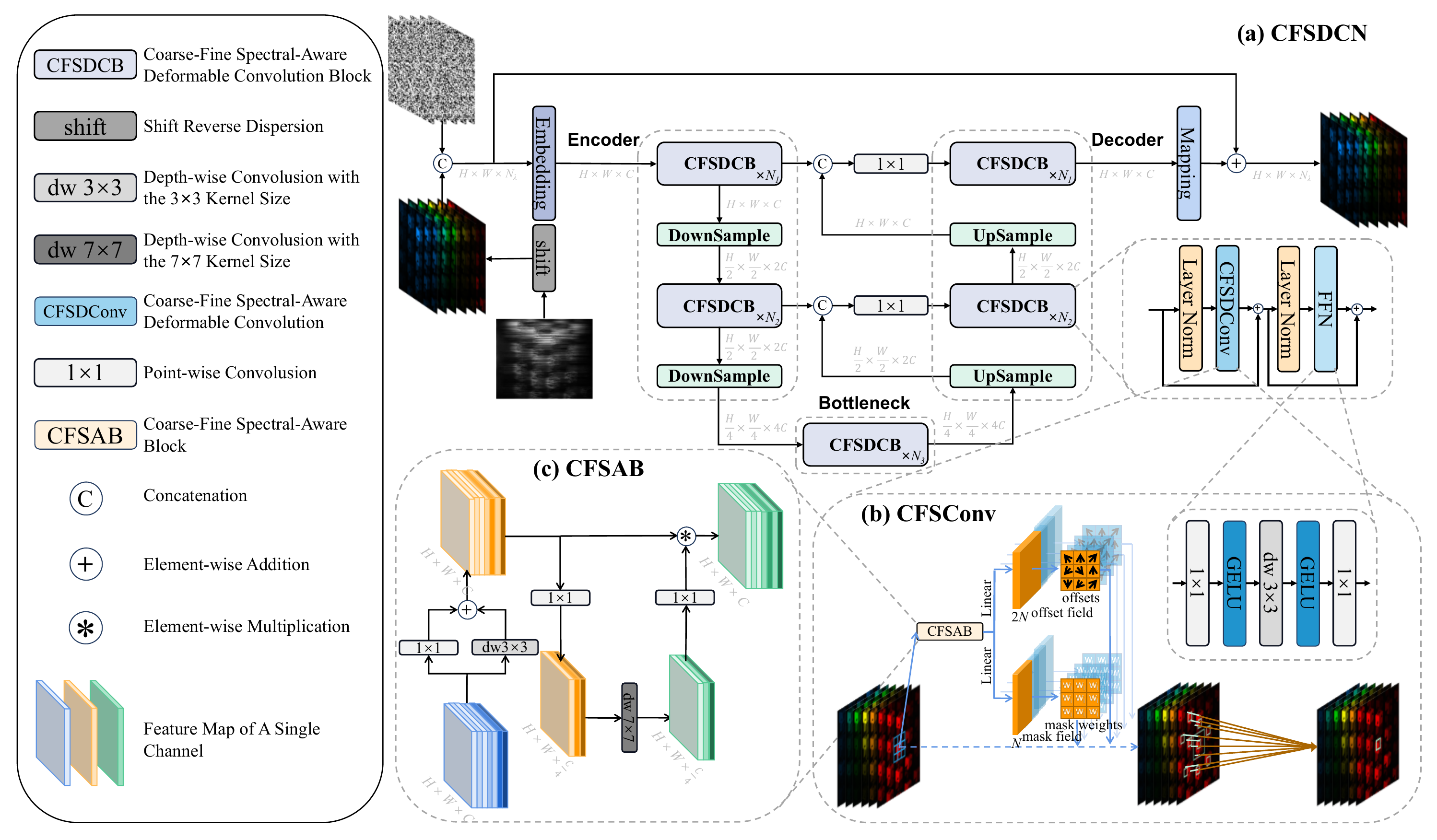} % 替换为你的图片路径
  \vspace{-3mm}
  \caption{Details of CFSDCN.    (a) Overall structure diagram of CFSDCN. (b) Schematic diagram of Coarse-Fine Spectral-Aware Deformable Convolution (CFSConv). (c) Schematic diagram of Coarse-Fine Spectral-Aware Block (CFSAB).} % 图片标题
  \label{fig:cfsdcn} % 用于引用的标签
  \vspace{-5mm}
\end{figure*}

The overall structure of CFSDCN is shown in the Fig.~\ref{fig:cfsdcn} (a). For a given 2D measurement $\mathbf{Y} \in \mathbb{R}^{H \times (W + d(N_{\lambda} - 1))}$, we reverse the dispersion process (Eq . (\ref{2})) and shift back the measurement to obtain the input signal $\mathbf{H} \in \mathbb{R}^{H \times W \times N_{\lambda}}$ as
\begin{equation}
      \mathbf {H}(x,y,n_\lambda ) = \mathbf {Y}(x,y-d(\lambda _n-\lambda _c)). \vspace {-1.2mm} \label {4}
\end{equation}
\textbf{Subsequently}, we concatenate $\mathbf{H}$ with the 3D mask $\mathbf{M} \in \mathbb{R}^{H \times W \times N_{\lambda} }$ and employ a \textit{conv }1$\times$1 (a convolution operation with a kernel size of 1) to obtain the initialized feature $\mathbf{X} \in \mathbb{R}^{H \times W \times N_{\lambda} }$ .
\textbf{Then}, referring to the Unet structure, the main architecture of our network consists of three parts: encoder, bottleneck, and decoder. Each layer of encoder consists of $N_1$ Coarse-Fine Spectral-Aware Deformable Convolution Blocks (CFSDCBs) and downsampling, while each layer of decoder consists of $N_2$ CFSDCBs and corresponding upsampling. The bottleneck contains $N_3$ CFSDCBs. The structure of CFSDCB is shown in Fig.~\ref{fig:cfsdcn} (a). \textbf{Finally}, a \textit{conv}3$\times$3 is used to change the output of decoder $\mathbf{X_{d}} \in \mathbb{R}^{H \times W \times C}$ to the residual HSIs $ \mathbf{R} \in \mathbb{R}^{H \times W \times N_{\lambda} }$. And the the reconstructed HSIs $X'$ can be obtained by the sum of $\mathbf{X}$ and $\mathbf{R}$, \textit{i.e.},$\mathbf{X'} = \mathbf{X} + \mathbf{R}$.

\subsection{Spatial-aware Deformable Convolution Module}
\textbf{Grouped Deformable Convolution.} In traditional Convolutional Neural Network (CNN) based approaches, effectively leveraging the inherent non-local similarity in hyperspectral images (HSI) poses a challenge. To solve this problem, we introduce deformable convolution, which captures non-local features and long-range dependencies well through an offset.Moreover, the 3×3 size convolution kernel is friendly in terms of speed and calculation amount, so our goal is to explore the potential of deformable convolution in HSI reconstruction. We introduce Grouped Deformable Convolution \cite{dcnv3} to extract spatially different feature information of the spectrum. Fig.~\ref{fig:cfsdcn} (b) showcases the structure of our deformable convolution used in our approach. For each pixel point $p_{0}$ of the input $\mathbf{X_{in}} \in \mathbb{R}^{H \times W \times C}$, the fundamental principle of our deformable convolution can be expressed as
\begin{equation}
    X(p{0}) = \textstyle \sum^{G}_{g=1}\sum^{K}_{k=1}w_{g}m_{gk}x_{g}(p_{0}+p_{k}+\Delta p_{gk}), \vspace {-1.2mm} \label {5} 
\end{equation}

where $G$ represents the total number of aggregation groups similar to the multi-head concept, $K$ indicates the total number of subsample points under the changed kernel size, and $w_{g}$ refers to the location-irrelevant projection weights of each group, $m_{gk}$ denotes the modulation scalar for the \textit{k}-th sampling point in the \textit{g}-th group, $x_{g}$ is the input feature mapping of the slice, and $\Delta p_{gk}$ is the offset of the \textit{k}-th sampling point in the \textit{g}-th group  relative to the original $p_{k}$ sampling point.

\begin{table*}[t]
    \renewcommand{\arraystretch}{1.0}
    \caption{Comparisons of Params, FLOPs, PSNR (upper entry in each cell), and
    SSIM (lower entry in each cell) of different methods on 10 simulation scenes
    (Scene1-Scene10). Best results are in bold and the second-best results are underlined.}
     % \vspace{-5mm}
    \newcommand{\tabincell}[2]{\begin{tabular}{@{}#1@{}}#2\end{tabular}}
    % \caption{SSIM values by different algorithms on 10 synthetic data.}
    \centering
    \resizebox{1.00\textwidth}{!}
    {
    \centering
    % \begin{tabular}{c|c|c|c|c|c|c|>{\columncolor{lightgray}}c}
    % \begin{tabular}{c|c|c|c|c|c|c|c|c|c|c|c}
    \begin{tabular}{cccccccccccccc}
% 
    % \hline
    \toprule
    Algorithms 
    & Params
    & GFLOPs
    & Scene1
    & Scene2
    & Scene3
    & Scene4
    & Scene5
    & Scene6
    & Scene7
    & Scene8
    & Scene9
    & Scene10
    & Average
    \\
    \hline
    % \midrule
    TwIST \cite{TwIST}
    &\tabincell{c}{-}
    &\tabincell{c}{-}
    &\tabincell{c}{25.16\\0.700} 
    &\tabincell{c}{23.02\\0.604} 
    &\tabincell{c}{21.40\\0.711} 
    &\tabincell{c}{30.19\\0.851} 
    &\tabincell{c}{21.41\\0.635} 
    &\tabincell{c}{20.95\\0.644} 
    &\tabincell{c}{22.20\\0.643} 
    &\tabincell{c}{21.82\\0.650} 
    &\tabincell{c}{22.42\\0.690} 
    &\tabincell{c}{22.67\\0.569} 
    &\tabincell{c}{23.12\\0.669}
    \\
    \hline
    GAP-TV \cite{GAPNet}
    &\tabincell{c}{-}
    &\tabincell{c}{-}
    &\tabincell{c}{26.82\\0.754} 
    &\tabincell{c}{22.89\\0.610} 
    &\tabincell{c}{26.31\\0.802} 
    &\tabincell{c}{30.65\\0.852} 
    &\tabincell{c}{23.64\\0.703} 
    &\tabincell{c}{21.85\\0.663} 
    &\tabincell{c}{23.76\\0.688} 
    &\tabincell{c}{21.98\\0.655} 
    &\tabincell{c}{22.63\\0.682} 
    &\tabincell{c}{23.10\\0.584} 
    &\tabincell{c}{24.36\\0.669}
    \\
    \hline
    DeSCI \cite{handcraft06}
    &\tabincell{c}{-}
    &\tabincell{c}{-}
    &\tabincell{c}{27.13\\0.748} 
    &\tabincell{c}{23.04\\0.620} 
    &\tabincell{c}{26.62\\0.818} 
    &\tabincell{c}{34.96\\0.897} 
    &\tabincell{c}{23.94\\0.706} 
    &\tabincell{c}{22.38\\0.683} 
    &\tabincell{c}{24.45\\0.743} 
    &\tabincell{c}{22.03\\0.673} 
    &\tabincell{c}{24.56\\0.732} 
    &\tabincell{c}{23.59\\0.587} 
    &\tabincell{c}{25.27\\0.721}
    \\
    \hline
    $\lambda$-net \cite{cnnsci04}
    &\tabincell{c}{62.64M}
    &\tabincell{c}{117.98}
    &\tabincell{c}{30.10\\0.849} 
    &\tabincell{c}{28.49\\0.805} 
    &\tabincell{c}{27.73\\0.870} 
    &\tabincell{c}{37.01\\0.934} 
    &\tabincell{c}{26.19\\0.817} 
    &\tabincell{c}{28.64\\0.853} 
    &\tabincell{c}{26.47\\0.806} 
    &\tabincell{c}{26.09\\0.831} 
    &\tabincell{c}{27.50\\0.826} 
    &\tabincell{c}{27.13\\0.816} 
    &\tabincell{c}{28.53\\0.841}
    \\
    \hline
    HSSP \cite{deepunf}
    &\tabincell{c}{-}
    &\tabincell{c}{-}
    &\tabincell{c}{31.48\\0.858} 
    &\tabincell{c}{31.09\\0.842} 
    &\tabincell{c}{28.96\\0.823} 
    &\tabincell{c}{34.56\\0.902} 
    &\tabincell{c}{28.53\\0.808} 
    &\tabincell{c}{30.83\\0.877} 
    &\tabincell{c}{28.71\\0.824} 
    &\tabincell{c}{30.09\\0.881} 
    &\tabincell{c}{30.43\\0.868} 
    &\tabincell{c}{28.78\\0.842} 
    &\tabincell{c}{30.35\\0.852}
    \\
    \hline
    DNU \cite{DNU}
    &\tabincell{c}{1.19M}
    &\tabincell{c}{163.48}
    &\tabincell{c}{31.72\\0.863} 
    &\tabincell{c}{31.13\\0.846} 
    &\tabincell{c}{29.99\\0.845} 
    &\tabincell{c}{35.34\\0.908} 
    &\tabincell{c}{29.03\\0.833} 
    &\tabincell{c}{30.87\\0.887} 
    &\tabincell{c}{28.99\\0.839} 
    &\tabincell{c}{30.13\\0.885} 
    &\tabincell{c}{31.03\\0.876} 
    &\tabincell{c}{29.14\\0.849} 
    &\tabincell{c}{30.74\\0.863}
    \\
    \hline
    DIP-HSI \cite{plug04}
    &\tabincell{c}{33.85M}
    &\tabincell{c}{64.42}
    &\tabincell{c}{32.68\\0.890} 
    &\tabincell{c}{27.26\\0.833} 
    &\tabincell{c}{31.30\\0.914} 
    &\tabincell{c}{40.54\\0.962} 
    &\tabincell{c}{29.79\\0.900} 
    &\tabincell{c}{30.39\\0.877} 
    &\tabincell{c}{28.18\\0.913} 
    &\tabincell{c}{29.44\\0.874}
    &\tabincell{c}{34.51\\0.927}
    &\tabincell{c}{28.51\\0.851}
    &\tabincell{c}{31.26\\0.894}
    \\
    \hline
    TSA-net \cite{TSA-Net}
    &\tabincell{c}{44.25M}
    &\tabincell{c}{110.06}
    &\tabincell{c}{32.03\\0.892} 
    &\tabincell{c}{31.00\\0.858} 
    &\tabincell{c}{32.25\\0.915} 
    &\tabincell{c}{39.19\\0.953} 
    &\tabincell{c}{29.39\\0.884} 
    &\tabincell{c}{31.44\\0.908} 
    &\tabincell{c}{30.32\\0.878} 
    &\tabincell{c}{29.35\\0.888} 
    &\tabincell{c}{30.01\\0.890} 
    &\tabincell{c}{29.59\\0.874} 
    &\tabincell{c}{31.46\\0.894}
    \\
    \hline
    DGSMP \cite{DGSMP}
    &\tabincell{c}{3.76M}
    &\tabincell{c}{646.65}
    &\tabincell{c}{33.26\\0.915} 
    &\tabincell{c}{32.09\\0.898} 
    &\tabincell{c}{33.06\\0.925} 
    &\tabincell{c}{40.54\\0.964} 
    &\tabincell{c}{28.86\\0.882} 
    &\tabincell{c}{33.08\\0.937} 
    &\tabincell{c}{30.74\\0.886} 
    &\tabincell{c}{31.55\\0.923} 
    &\tabincell{c}{31.66\\0.911} 
    &\tabincell{c}{31.44\\0.925} 
    &\tabincell{c}{32.63\\0.917}
    \\
    \hline
    HDNet \cite{cnnsci03}
    &\tabincell{c}{2.37M}
    &\tabincell{c}{154.76}
    &\tabincell{c}{35.14\\0.935} 
    &\tabincell{c}{35.67\\0.940} 
    &\tabincell{c}{36.03\\0.943} 
    &\tabincell{c}{42.30\\0.969} 
    &\tabincell{c}{32.69\\0.946} 
    &\tabincell{c}{34.46\\0.952} 
    &\tabincell{c}{33.67\\0.926} 
    &\tabincell{c}{32.48\\0.941} 
    &\tabincell{c}{34.89\\0.942} 
    &\tabincell{c}{32.38\\0.937} 
    &\tabincell{c}{34.97\\0.943}
    \\
    \hline
    MST-S \cite{MST}
    &\tabincell{c}{0.93M}
    &\tabincell{c}{12.96}
    &\tabincell{c}{34.71\\0.930} 
    &\tabincell{c}{34.45\\0.925} 
    &\tabincell{c}{35.32\\0.943} 
    &\tabincell{c}{41.50\\0.967} 
    &\tabincell{c}{31.90\\0.933} 
    &\tabincell{c}{33.85\\0.943} 
    &\tabincell{c}{32.69\\0.911} 
    &\tabincell{c}{31.69\\0.933} 
    &\tabincell{c}{34.67\\0.939} 
    &\tabincell{c}{31.82\\0.926} 
    &\tabincell{c}{34.26\\0.935}
    \\
    \hline
    MST-M \cite{MST}
    &\tabincell{c}{1.50M}
    &\tabincell{c}{18.07}
    &\tabincell{c}{35.15\\0.937} 
    &\tabincell{c}{35.19\\0.935} 
    &\tabincell{c}{36.26\\0.950} 
    &\tabincell{c}{42.48\\0.973} 
    &\tabincell{c}{32.49\\0.943} 
    &\tabincell{c}{34.28\\0.948} 
    &\tabincell{c}{33.29\\0.921} 
    &\tabincell{c}{32.40\\0.943} 
    &\tabincell{c}{35.35\\0.942} 
    &\tabincell{c}{32.53\\0.935} 
    &\tabincell{c}{34.94\\0.943}
    \\
    \hline
    MST-L \cite{MST}
    &\tabincell{c}{2.03M}
    &\tabincell{c}{28.15}
    &\tabincell{c}{35.40\\0.941} 
    &\tabincell{c}{35.87\\0.944} 
    &\tabincell{c}{36.51\\0.953} 
    &\tabincell{c}{42.27\\0.973} 
    &\tabincell{c}{32.77\\0.947} 
    &\tabincell{c}{34.80\\0.955} 
    &\tabincell{c}{33.66\\0.925} 
    &\tabincell{c}{32.67\\0.948} 
    &\tabincell{c}{35.39\\0.949} 
    &\tabincell{c}{32.50\\0.941} 
    &\tabincell{c}{35.18\\0.948}
    \\
    \hline
    CST-S \cite{CST}
    &\tabincell{c}{1.20M}
    &\tabincell{c}{11.67}
    &\tabincell{c}{34.78\\0.930}
    &\tabincell{c}{34.81\\0.931}
    &\tabincell{c}{35.42\\0.944}
    &\tabincell{c}{41.84\\0.967}
    &\tabincell{c}{32.29\\0.939}    
    &\tabincell{c}{34.49\\0.949}
    &\tabincell{c}{33.47\\0.922}
    &\tabincell{c}{32.89\\0.945}
    &\tabincell{c}{34.96\\0.944}
    &\tabincell{c}{32.14\\0.932}
    &\tabincell{c}{34.71\\0.940}
    \\
    \hline
    CST-M \cite{CST}
    &\tabincell{c}{1.36M}
    &\tabincell{c}{16.91}
    &\tabincell{c}{35.16\\0.938} 
    &\tabincell{c}{35.60\\0.942} 
    &\tabincell{c}{36.57\\0.953} 
    &\tabincell{c}{42.29\\0.972} 
    &\tabincell{c}{32.82\\0.948} 
    &\tabincell{c}{35.15\\0.956} 
    &\tabincell{c}{33.85\\0.927} 
    &\tabincell{c}{33.52\\0.952} 
    &\tabincell{c}{35.28\\0.946} 
    &\tabincell{c}{32.84\\0.940} 
    &\tabincell{c}{35.31\\0.947}
    \\
    \hline
    
    CST-L \cite{CST}
    &\tabincell{c}{3.00M}
    &\tabincell{c}{40.10}
    &\tabincell{c}{\textbf{35.96}\\\underline{0.949}} 
    &\tabincell{c}{\underline{36.84}\\\underline{0.955}} 
    &\tabincell{c}{\underline{38.16}\\0.962} 
    &\tabincell{c}{42.44\\0.975} 
    &\tabincell{c}{33.25\\0.955} 
    &\tabincell{c}{\textbf{35.72}\\0.963} 
    &\tabincell{c}{\underline{34.86}\\\underline{0.944}} 
    &\tabincell{c}{\textbf{34.34}\\\underline{0.961}} 
    &\tabincell{c}{36.51\\0.957} 
    &\tabincell{c}{\textbf{33.09}\\0.945} 
    &\tabincell{c}{\underline{36.12}\\0.957}
    \\
    \hline
    % \rowcolor{lightgray}
    \bf{CFSDCN-S (Ours)}
    &\tabincell{c}{\textbf{0.76M}}
    &\tabincell{c}{\textbf{9.45}}
    &\tabincell{c}{35.45\\0.944} 
    &\tabincell{c}{35.88\\0.944} 
    &\tabincell{c}{37.00\\0.958} 
    &\tabincell{c}{42.94\\0.978} 
    &\tabincell{c}{32.99\\0.950} 
    &\tabincell{c}{35.17\\0.959} 
    &\tabincell{c}{34.12\\0.936} 
    &\tabincell{c}{33.17\\0.954} 
    &\tabincell{c}{36.52\\0.959} 
    &\tabincell{c}{32.49\\0.944} 
    &\tabincell{c}{35.57\\0.953}
    \\
    \hline
    % \rowcolor{lightgray}
    \bf{CFSDCN-M (Ours)}
    &\tabincell{c}{1.7M}
    &\tabincell{c}{18.1}
    &\tabincell{c}{35.59\\0.947} 
    &\tabincell{c}{36.58\\0.954} 
    &\tabincell{c}{37.94\\\underline{0.963}} 
    &\tabincell{c}{\underline{43.33}\\\underline{0.982}} 
    &\tabincell{c}{\underline{33.30}\\\underline{0.956}} 
    &\tabincell{c}{35.38\\\underline{0.964}} 
    &\tabincell{c}{34.36\\0.939} 
    &\tabincell{c}{33.46\\0.959} 
    &\tabincell{c}{\underline{37.17}\\\underline{0.964}} 
    &\tabincell{c}{\underline{32.97}\\\underline{0.948}} 
    &\tabincell{c}{36.01\\\underline{0.958}} 
    \\
    \hline
    \bf{CFSDCN-L (Ours)}
    &\tabincell{c}{2.52M}
    &\tabincell{c}{31.0}
    &\tabincell{c}{\underline{35.93}\\\textbf{0.950}} 
    &\tabincell{c}{\textbf{37.06}\\\textbf{0.958}} 
    &\tabincell{c}{\textbf{38.58}\\\textbf{0.970}} 
    &\tabincell{c}{\textbf{43.80}\\\textbf{0.983}} 
    &\tabincell{c}{\textbf{33.38}\\\textbf{0.957}} 
    &\tabincell{c}{\underline{35.52}\\\textbf{0.966}} 
    &\tabincell{c}{\textbf{34.88}\\\textbf{0.946}} 
    &\tabincell{c}{\underline{33.51}\\\textbf{0.962}} 
    &\tabincell{c}{\textbf{37.50}\\\textbf{0.965}} 
    &\tabincell{c}{32.92\\\textbf{0.950}} 
    &\tabincell{c}{\textbf{36.31}\\\textbf{0.961}} 

    \\
    \bottomrule
    \end{tabular}
    }
    \vspace{-5mm}
    \label{Tab:simu}
\end{table*}

\subsection{Coarse-Fine Spectral-Aware Module}
In the proposal and application of deformable convolutions, the similarity and attention across channels have not been adequately emphasized and utilized, which is a crucial property for HSIs. Therefore, we explore CNN-based spectral perception methods to effectively utilize this aspect of HSIs.

\textbf{Fine-grained spectral similarity.} We note that in DCN, to reduce the number of parameters, they employ depth-wise separable convolutions instead of conventional convolution operations. This approach does not allow the weights of offsets and masks to consider the spectral similarity at the same location across different spectra brought about by the eventual point-wise convolution, which we refer to as fine-grained spectral similarity. Therefore, as shown in the left part of Fig.~\ref{fig:cfsdcn} (c), we add a parallel branch to learn this fine-grained spectral similarity
\begin{gather}
    \mathbf{X_{\text{p}}} = \text{Conv}_{1\times1}(\mathbf{X_{\text{inp}}}), \\
    \mathbf{X_{\text{fine}}} = \text{DWConv}_{3\times3}(\mathbf{X_{\text{inp}}}) + \mathbf{X_{\text{sconv}}},
\end{gather}
where $\mathbf{X_{\text{inp}}} \in \mathbb{R}^{H \times W \times C}$ represents the input of the CFSConv, $\mathbf{X_{\text{p}}} \in \mathbb{R}^{H \times W \times C}$ denotes the added parallel branch for capturing fine-grained spectral similarities, $\mathbf{X_{fine}} \in \mathbb{R}^{H \times W \times C}$ denotes the result of fine-gained spectral aware. We employ a point-wise convolution to facilitate interaction between information at the same spatial positions but across different spectral bands, and then integrate this output back into the original branch. This integration enables the model to acquire fine-grained spectral information.

\begin{figure*}[t]
  \centering
  \includegraphics[width=\textwidth]{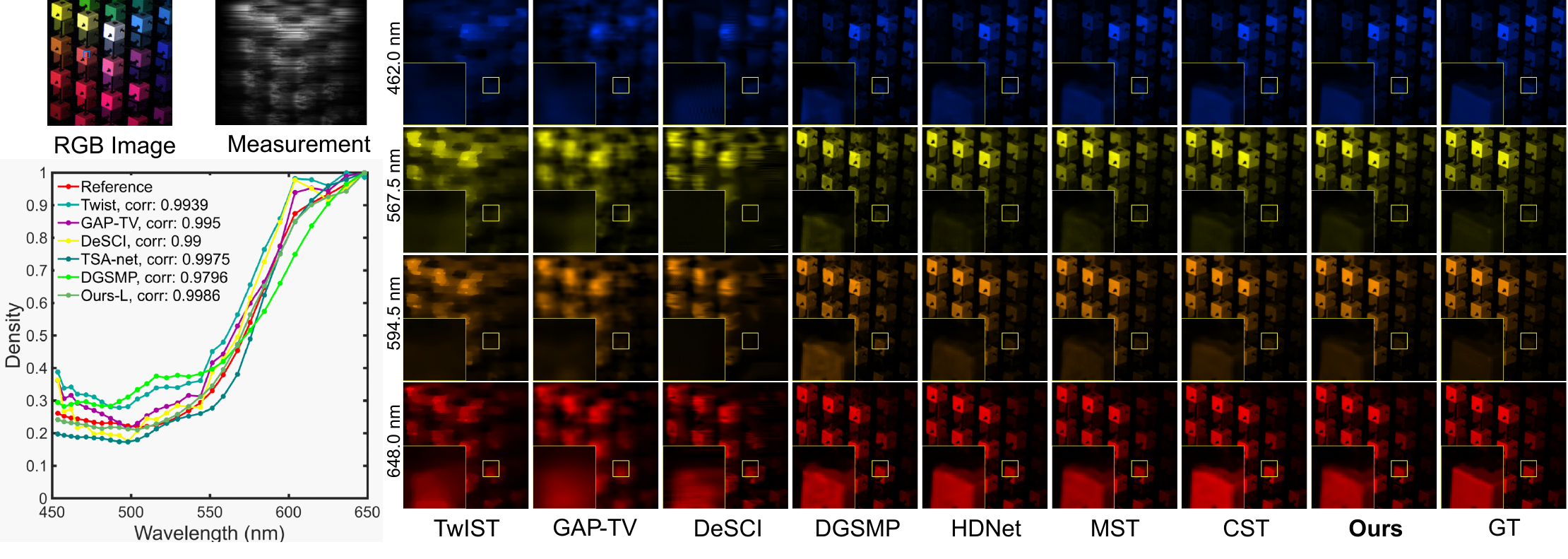} % 替换为你的图片路径
  \vspace{-5mm}
  \caption{Comparison of reconstruction results of simulation HSI datasets for 4 out of 28 spectral channels in Scene 2. 7 SOTA methods and our CFSDCN are included here. Zoom in for a better view.} % 图片标题
  \vspace{-4mm}
  \label{fig:sim} % 用于引用的标签
\end{figure*}

\textbf{Coarse-grained spectral similarity.} For coarse-grained spectral similarity, that is, the similarity between different locations across different spectral channels, we design a Large-kernel-guided Coarse-grained Spectral-aware(LCS) module. As shown in the right part of Fig.~\ref{fig:cfsdcn} (c), we use large kernel convolutions to increase the receptive field, capturing information between different positions, and then place it within a spectral information interaction module composed of point-wise convolutions
\begin{eqnarray}
    \mathbf{X_\text{down}} = {\rm Conv}_{1\times1}(\mathbf{X_\text{fine}}), \\
    \mathbf{X_\text{lkconv}} = {\rm DWConv}_{7\times7}(\mathbf{X_\text{down}}), \\
    \mathbf{X_\text{weight}} = {\rm Conv}_{1\times1}(\mathbf{X_\text{Coarse}}), 
\end{eqnarray}
where ${\mathbf{X}}_{down} \in \mathbb{R}^{H \times W \times C/4}$ refers to the result of $X_{fine}$ after dimensionality reduction through point-wise convolution, representing the process of channel information fusion; 
${\mathbf{X}}_{lkconv} \in \mathbb{R}^{H \times W \times C/4}$ indicates the coarse-grained information matrix obtained by gathering information around each pixel point via a depth-wise convolution with a large kernel; ${\mathbf{X}_{weight}} \in \mathbb{R}^{H \times W \times C}$
is a weight matrix obtained by amplifying the dimension by point-wise convolution, where each value represents the weighted information post coarse-grained spectral perception. Then multiplying it with the input ensures that each value in the input carries the coarse-grained spectral similarity features brought about by the large receptive field of the large kernel convolution
\begin{equation}
   \mathbf{X_\text{coarse}} = \mathbf{X_\text{weight}} \cdot \mathbf{X_\text{fine}},
\end{equation}
where $\mathbf{X_{coarse}} \in \mathbb{R}^{H \times W \times C}$ denotes the results obtained after coarse-grained spectral perception.

\section{EXPERIMENTS}
\label{sec:EXPERIMENTS}

\subsection{Experimental Settings}
Consistent with some previous studies \cite{TSA-Net,MST,CST,DAUHST}, we select 28 bands ranging from 450nm to 650nm for spectral interpolation to obtain data for experimentation. We conduct experiments on both simulated and real HSI datasets. 

\vspace{1mm}
\noindent\textbf{Simulation HSI Data.} For the simulation, we engage the CAVE \cite{CAVE} and KAIST \cite{KAIST} datasets, both widely used in the field. The CAVE dataset contains 32 hyperspectral images with dimensions of 256$\times$256, and the KAIST dataset includes 30 hyperspectral images, each measuring 2704$\times$3376. Adhering to the methodology established by TSA-Net, we designate CAVE for training and select 10 scenes from KAIST for evaluation purposes.

\vspace{1mm}
\noindent\textbf{Real Dataset.} Inspired by the approach used in previous works, we conduct tests using a real HSI dataset acquired by the CASSI system to the generalization of our model. We use the real HSI dataset collected by the CASSI system developed in TSA-Net \cite{TSA-Net}.

\begin{figure*}[t]
  \centering
  \includegraphics[width=\textwidth]{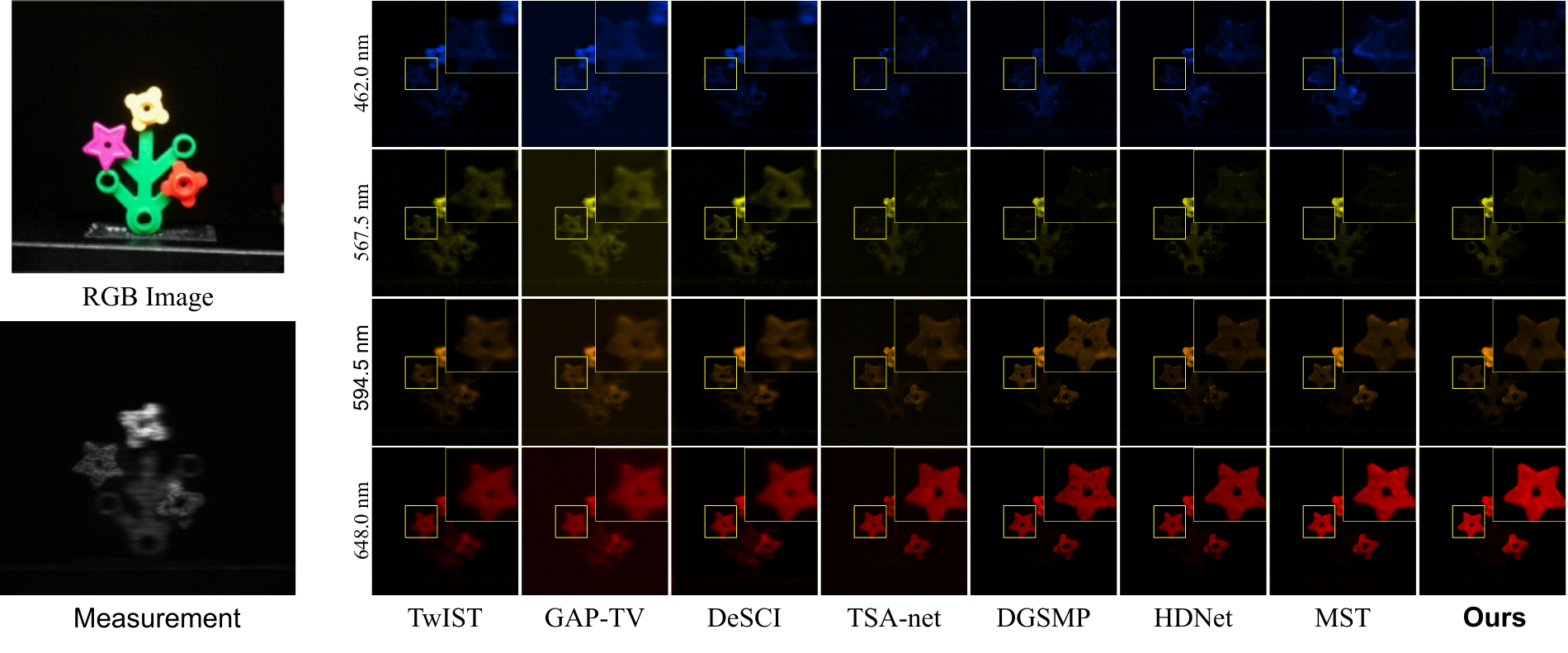} % 替换为你的图片路径
  \caption{Comparison of reconstruction results of real HSI datasets for 4 out of 28 spectral channels in Scene 2. 7 SOTA methods and our CFSDCN-M are included here. Zoom in for a better view.} % 图片标题
  \label{fig:real} % 用于引用的标签
\vspace{2mm}
\end{figure*}

\vspace{1mm}
\noindent\textbf{Evaluation Metrics.} The performance of HSI reconstruction is evaluated by Peak Signal-to-Noise Ratio (PSNR) and Structural Similarity Index (SSIM) \cite{SSIM}.

\vspace{1mm}
\noindent\textbf{Implementation Details.} The CFSDCN model, developed using PyTorch, utilizes the Adam optimizer with hyperparameters $\beta_1$ = 0.9 and $\beta_2$ = 0.999. Training spanned 500 epochs, with the learning rate initially set at $4 \times 10^{-4}$ and reduced by half every 50 epochs. Training samples were generated by randomly cropping 256×256 cubes from a 28-channel HSI cube for simulated data, and 660×660 cubes for actual experimental data. The model employed a dispersion step size of 2 and a batch size of 5, incorporating random rotations and flips for data augmentation. The training and evaluation were performed on a single RTX 3090 GPU.

\subsection{Quantitative Results}
In our study, we comprehensively compare the CFSDCN method with several current end-to-end methods including model-based methods \cite{TwIST,handcraft05,handcraft06}, deep CNN-based methods \cite{deepunf,DNU,TSA-Net,cnnsci04,plug04,DGSMP,cnnsci03}, and Transformer-based methods\cite{MST,CST}.
To ensure consistency, we re-train all methods using the same training set and tested them using the same settings as in DGSMP \cite{DGSMP}. The PSNR and SSIM results for these methods across ten scenarios in the test set are shown in Tab.~\ref{Tab:simu}. The parameters (Params) and floating-point operations (FLOPs) of each method are also given in the table. From these tables, we can observe that our method has significant improvements in reconstruction performance and computational cost.

\begin{table*}[htbp]
\centering
\caption{Combined Table of Spatial-Spectral-Aware Mechanism, and CNN-based Spectral-Aware Mechanism Comparisons.}
\label{tab:combined}
% 第一个子表格
\begin{subtable}{.45\textwidth}
\centering
\caption{Spatial-Spectral-Aware Mechanism Comparison.}
\label{subtab:SA}
\resizebox{1.3\textwidth}{!}{
\begin{tabular}{ccccccccc}
\toprule
Method & Baseline & Conv & G-MSA & W-MSA & Swin-MSA & S-MSA & SAH-MSA & \textbf{CFSDCN} \\
\midrule
PSNR      & 32.57 &  33.63 & 35.04 & 35.02 & 35.12 & 35.21 & 35.53 & \textbf{35.57} \\
SSIM      & 0.906 &  0.931 & 0.944 & 0.943 & 0.945 & 0.946 & 0.948 & \textbf{0.953} \\
Params(M) & 0.51  &  0.52  & 1.85  & 1.85  & 1.85  & 1.66  & 1.36  & \textbf{0.759}  \\
FLOPs(G)  & 6.40  &  6.69 & 35.31 & 24.98 & 24.98  & 24.74 & 24.60 & \textbf{9.45} \\
\bottomrule
\end{tabular}
}
\end{subtable}
\hfill % 这个命令在子表格之间添加一些空间
% 第三个子表格
\begin{subtable}{.45\textwidth}
\centering
\caption{CNN-based Spectral-Aware Mechanism Comparison.}
\label{subtab:fined}
\resizebox{0.8\textwidth}{!}{
\begin{tabular}{cccccc}
\toprule
Method & SEBlock & Conv3$\times$3 & \textbf{Conv7$\times$7} & Conv11$\times$11\\
\midrule
PSNR      & 35.37          &  35.51 & \textbf{35.57} & 35.56 \\
SSIM      & 0.951          &  0.952 & \textbf{0.953} & 0.952 \\
Params(M) & \textbf{0.758} &  0.760 & 0.764          & 0.774 \\
FLOPs(G)  & \textbf{9.13}  &  9.37 &  9.45           & 9.60  \\
\bottomrule
\end{tabular}
}
\end{subtable}
% \vspace{-4mm}
\end{table*}

\begin{table}[htbp]
\centering
\caption{Break-down Ablation.}
\label{tab:breakdown}
\resizebox{.48\textwidth}{!}{
\begin{tabular}{cccccccc}
\toprule
BaseLine & DCB & CFSAB & PSNR & SSIM & Params (M) & FLOPs(G) \\
\midrule
\checkmark &            &            & 33.63 & 0.931 & \textbf{0.52} & \textbf{6.69} \\
\checkmark & \checkmark &            & 35.28 & 0.949 & 0.76 & 8.71 \\
\checkmark & \checkmark & \checkmark & \textbf{35.57} & \textbf{0.953} & 0.76 & 9.45 \\
\bottomrule
\end{tabular}
}
\vspace{-3mm}
\end{table}

\subsection{Qualitative Results}
\textbf{Simulation HSI Reconstruction.} 
Fig.~\ref{fig:sim} demonstrates the simulated HSI reconstruction performance of our CFSDCN and other SOTA methods in four spectral channels of Scene 2. We have magnified the images within the yellow box and placed them in the lower-left corner of the picture for a detailed comparison.
It can be seen that, compared with over-smoothing, loss of fine-grained structure, or edge distortion and some unwanted color artifacts in other images, our CFSDCN is more effective in reconstructing sharp edge areas, achieving perceptually pleasing image while maintaining uniform spatial smoothness.
Furthermore, we plot the spectral density curve of the reconstructed area versus the ground truth (lower left) to demonstrate the effectiveness of our proposed CFSDCN in achieving consistent spectral reconstruction.

\noindent\textbf{Real HSI Reconstruction.} We then apply our proposed method to real HSI reconstruction. Following the settings in \cite{DGSMP,MST} to ensure fairness, we introduce 11-bit shot noise in the training process to simulate real measurement conditions. As shown in Fig.~\ref{fig:real}, we compare our method with seven SOTA methods for real scenes. In Scene 1, the reconstruction results obtained from our method demonstrate a smoother appearance, more detailed edge processing, and visually pleasing effects. These outcomes attest to the reliability, robustness, and enhanced generalization capabilities of our approach.

\subsection{Ablation Study}
\noindent{\bf Break-down Ablation.} We conduct a break-down ablation experiment to study the impact of each structure on achieving higher performance. The results are shown in Tab.~\ref{tab:breakdown}. These results sufficiently demonstrate the effectiveness of both DCB and CFSAB. 

\noindent{\bf Spatial-Spectral-Aware Mechanism Comparison.} We compare our CFSDCB with other spatial-spectral-aware mechanisms. These results are showed in Tab.~\ref{tab:combined} (a). To ensure fairness, we select several models with similar effects for comparison. This result highlights the computational efficiency of our method, mainly because our fully convolutional structure is much more economical than computing self-attention, significantly reducing the computational effort.

\noindent{\bf {CNN-based Spectral-Aware Mechanism Comparison.}} We primarily compare the effects of SEBlock \cite{seblock}, \textit{conv}3$\times$3 coarse-grained spectral mechanism, and our LCS (use \textit{conv}7$\times$7), with results shown in Tab.~\ref{tab:combined} (b). It is clear that our LCS successfully combines the advantages of these approaches, choosing an appropriately sized convolution kernel to obtain large improvements at a very low overhead.

\section{CONCLUSION}
\label{sec:CONCLUSION}
In this paper, we present a novel CNN-based method for hyperspectral image (HSI) reconstruction, CFSDCN. Inspired by the inter-spectral characteristics of HSIs and the latest advancements in deformable convolutions, we propose using grouped deformable convolutions to capture spatial features in HSIs. Additionally, we design a convolution-based coarse-fine granularity spectral perception module, which enables the offsets and masks of deformable convolutions to take into account similar information between spectra. Based on this, we have developed a series of highly effective CFSDCN models. Quantitative experiments show that our model significantly outperforms the SOTA algorithm and requires lower computational cost 
 and parameter amount. Qualitative experiments show that our CFSDCN can reconstruct more visually pleasing HSIs.

\noindent{\bf{Acknowledgements}}
This work is partially conducted when Jincheng Yang was a summer intern in Sensing and Computational Imaging (SCI) Lab in 2023.
This work was supported by the National Natural Science Foundation of China (grant number 62271414), Zhejiang Provincial Outstanding Youth Science Foundation (grant number LR23F010001) and the Key
Project of Westlake Institute for Optoelectronics (grant number 2023GD007).

\bibliographystyle{IEEEbib}
\small
\bibliography{strings,refs}

\end{document}